\documentclass[fleqn,10pt]{wlscirep}
\usepackage[utf8]{inputenc}
\usepackage[T1]{fontenc}
\usepackage{booktabs,multirow,xcolor, amsmath}

\def\yv{\boldsymbol y}
\def\epsilonv{\boldsymbol \epsilon}

\def\Uv{\boldsymbol U}
\def\Vv{\boldsymbol V}
\def\Wv{\boldsymbol W}
\def\Xv{\boldsymbol X}
\def\Yv{\boldsymbol Y}
\newcommand{\Sigmav}{\mbox{\boldmath{$\Sigma$}}}
\def\Xbf{\mathbf X}

\def\1v{\mathbf 1}

\title{Detection of LUAD-Associated Genes Using Wasserstein Distance in Multi-Omics Feature Selection}

\author[1,*]{Shaofei Zhao}
\author[2, +]{Siming Huang}
\author[3, +]{Kexuan Li}
\author[4, +]{Weiyu Zhou}
\author[1,2,+]{Lingli Yang}
\author[2,+]{Shige Wang}
\affil[1]{Binghamton University}
\affil[2]{Affiliation, department, city, postcode, country}

\affil[*]{szhao38@binghamton.edu}

\affil[+]{these authors contributed equally to this work}

\begin{abstract}
Lung adenocarcinoma (LUAD) is characterized by substantial genetic heterogeneity, posing challenges in identifying reliable biomarkers for improved diagnosis and treatment. Tumor Mutational Burden (TMB) has traditionally been regarded as a predictive biomarker, given its association with immune response and treatment efficacy. In this study, we treated TMB as a response variable to identify genes highly correlated with it, aiming to understand its genetic drivers. We conducted a thorough investigation of recent feature selection methods through extensive simulations, selecting PC-Screen, DC-SIS, and WD-Screen as top performers. These methods handle multi-omics structures effectively, and can accommodate both categorical and continuous data types at the same time for each gene. Using data from The Cancer Genome Atlas (TCGA) via cBioPortal, we combined copy number alteration (CNA), mRNA expression and DNA methylation data as multi-omics predictors and applied these methods, selecting genes consistently identified across all three methods. 13 common genes were identified, including \emph{HSD17B4, PCBD2}, which show strong associations with TMB. Our multi-omics strategy and robust feature selection approach provide insights into the genetic determinants of TMB, with implications for targeted LUAD therapies.
\end{abstract}
\begin{document}

\flushbottom
\maketitle
%
%
\thispagestyle{empty}

\section*{Introduction}

Lung adenocarcinoma (LUAD), a predominant subtype of non-small cell lung cancer (NSCLC), accounts for nearly 40\% of all lung cancer cases worldwide, making it a critical focus in oncology research. According to the Global Cancer Observatory (GLOBOCAN), lung cancer remains the leading cause of cancer-related deaths, with over 2 million new cases and approximately 1.8 million deaths reported annually\cite{GlobalCancer}, and LUAD comprises the majority of these cases. The prognosis for LUAD patients remains poor, with a five-year survival rate below 20\%, particularly due to late-stage diagnoses when treatment options are limited.

Recent advancements in multi-omics technologies have provided a deeper understanding of the molecular landscape of LUAD, and several studies have focused on multi-omics approaches and machine learning techniques to extract highly related genes. For example, using feature selection frameworks with mutual information and random forest, the researchers found a consensus set of twelve genes with significant diagnostic potential, which could differentiate LUAD from normal samples with high accuracy\cite{abdelwahab2022feature}.However, despite these advancements, high genetic and molecular heterogeneity in LUAD continues to limit the applicability of existing biomarkers for early and personalized diagnostics.

The analysis of multi-omics data presents significant challenges due to its high dimensionality, with tens of thousands of genes but only a few hundred subjects, and the integration of multiple data platforms. This structure introduces a high degree of noise, and creates an array-like predictor structure, where for $n$ subjects, the predictors not only has the regular high dimensionality $p$ but also additional dimension $d$ associated with $d$ different platforms. This complexity makes traditional analysis methods ineffective and sometimes even incapable of handling predictors with such multi-layer structure. As a result, effective feature selection is crucial for detecting signals in this noisy environment.

Feature selection, or feature screening has long been a hot topic in statistical and machine learning research, particularly due to its critical role in managing high-dimensional and complex data. Fan and Lv's introduction of Sure Independent Screening (SIS) was a pivotal development, demonstrating that SIS could reliably identify all true predictors in sufficiently large samples, hence the term "sure" screening\cite{fan2008sure}. Their work inspired further advancements in sure screening, with methods like Distance Correlation based Sure Independence Screening (DC-SIS)\cite{li2012feature}, Projection based Sure Independence Screening (PC-Screen)\cite{liu2020model}, Stable Correlation Screening (SC-SIS)\cite{guo2022stable}, and Multivariate Rank Distance Correlation based Sure Independence Screening (MrDc-SIS)\cite{zhao2022distribution}. For a comprehensive review and comparison of these robust screening techniques, Zhao and Fu\cite{zhao2022distribution} offers detailed insights. Recently, Zhao \textit{et al.}\cite{zhao2024model} utilized a model-free and distribution-free screening method, MrDcGene, on the TCGA-LUAD dataset. This method effectively confirmed known biomarkers and identified new gene candidates associated with LUAD, showcasing the potential of advanced sure screening methods in handling the intrinsic complexity of multi-omics data.

Building on recent advancements in dependence measures, we noted a new approach using the Wasserstein distance as a dependence metric, as introduced in the work by Mordant and Segers\cite{mordant2022measuring}. The Wasserstein distance offers several advantages. First, it remains stable under transformations like rotation and monotonic changes, making it robust in noisy, high-dimensional data. Second, unlike some traditional measures, it does not rely on assumptions about linearity or Gaussian distributions, which allows it to work effectively in model-free, distribution-free settings, making it a versatile choice for complex, mulit-omics data. Finally, Wasserstein distance measures dependence by minimizing the ``transport cost'' between distributions, grounded in optimal transport theory, providing a reliable and theoretically sound way to measure associations. 

These properties make the Wasserstein distance particularly well-suited for high-dimensional, multi-omics datasets like the TCGA-LUAD, where both complexity and noise are prevalent. In this paper, we aim to thoroughly test and compare the performance of the Wasserstein distance against other feature selection methods through extensive simulation studies. Following these simulations, we will apply the Wasserstein distance-based approach to the real TCGA-LUAD dataset, which we downloaded through cBioPortal \cite{cerami2012cbio}, to evaluate its effectiveness in identifying meaningful gene associations for LUAD.

\section*{Results}

Up to three levels of \textbf{subheading} are permitted. Subheadings should not be numbered.

\subsection*{Comparative Analysis of Feature Selection Methods: Simulation Studies}
In our comparison, we evaluate ten popular feature selection methods, including our Wasserstein distance-based screening (WD-Screen). The other methods are as follows, for details, please refer to Zhao and Fu\cite{zhao2022distribution}
\begin{itemize}
    \item Sure Independence Screening (SIS)\cite{fan2008sure} -- uses Pearson correlation as the dependence measure between each predictor and response variable.
    \item  Sure Independence Ranking and Screening (SIRS)\cite{zhu2011model} -- uses Pearson correlation between each predictor and the rank of response variable.
    \item Robust Rank Correlation Screening (RRCS)\cite{li2012robust} -- uses Kendall's $\tau$ as the dependence measure between each predictor and the response variable.
    \item Distance Correlation based Sure Independence Screening (DC-SIS)\cite{li2012feature} -- uses distance correlation as the dependence measure between each predictor and the response variable.
    \item Robust Distance Correlation Screening (DC-RoSIS)\cite{zhong2016regularized} -- uses distance correlation between each predictor and the rank of response variable.
    \item Multivariate Rank based Distance Correlation Screening (MrDc-SIS)\cite{zhao2022distribution} -- uses distance correlation between the multivariate rank of predictors and response variables.
    \item Stable Correlation based Screening (SC-SIS)\cite{guo2022stable} -- uses a different weight function in the distance correlation as the dependence measure between each predictor and response variable.
    \item Projection Correlation based Screening (PC-Screen)\cite{liu2020model} -- uses projection correlation as the dependence measure between each predictor and response variable.
    \item Ball correlation based Screening (BCor-SIS)\cite{pan2018generic} -- uses ball correlation as the dependence measure between each predictor and response variable.
\end{itemize}
\subsubsection*{Study 1: Benchmarking and Validation of Feature Selection Methods}
To establish a benchmark, allow comparison with prior studies, and verify the correctness of our implementation for each method, we replicate a similar simulation setup as used in previous papers\cite{fan2008sure, li2012feature, zhao2022distribution}. In this study, we generate data with $n = 200$ observations and $p = 2000$ predictors. The predictors $\Xv_{n\times p}$ are drawn from a multivariate normal distribution with zero mean and AR (1) covariance structure, where the covariance matrix $\Sigma_{p\times p} = [\sigma_{ij}]_{p\times p}$, and $\sigma_{ij} = 0.5^{|i-j|}$ for $i, j = 1, 2, \dots, p$. The response variable $\yv_{n\times 1}$ is constructed by 
\[
\yv = \beta_1 \Xv_1 + \beta_2 \Xv_2 + \beta_3 \Xv_{12} + \beta_4 \Xv_{13} + \epsilonv,
\]
where $\epsilonv_{n\times 1}$ is standard normal and $\beta_{1,2,3,4} \sim Uniform (2, 5)$.

We repeat the simulation 200 times and follow Li \textit{et al.}\cite{li2012feature}, we utilize three criteria to assess the performance of the feature selection methods:
\begin{itemize}
    \item $\mathit{S}$: The minimum model size to include all true predictors. We draw a box plot of $\mathit{S}$. The smaller the $\mathit{S}$, the better the performance.
    \item $\mathit{P}_s$: The individual success rate of selecting a single true predictor within a predetermined cutoff across 200 replicates. The larger the $\mathit{P}_s$, the better the performance.
    \item $\mathit{P}_a$: The success rate of selecting all true predictors within a predetermined cutoff across 200 replicates. The larger the $\mathit{P}_a$, the better the performance.
\end{itemize}

We set the predetermined cutoff as $[n/\log(n)]$, where $[a]$ stands for the integer part of $a$, to be consistent with Li \textit{et al.}\cite{li2012feature}. For $n = 200$, the cutoff is $s = [n/\log(n)] = 37$. Besides the 3 criteria above, we also create a box plot for the rank of each true predictor. As a smaller rank indicates a more important predictor, ideally, the true predictors should be ranked at the top. This visualization can further highlight each method's effectiveness in identifying true predictors.

\begin{figure}[h!]
\centering
\begin{minipage}{.48\textwidth}
\includegraphics[width=\textwidth]{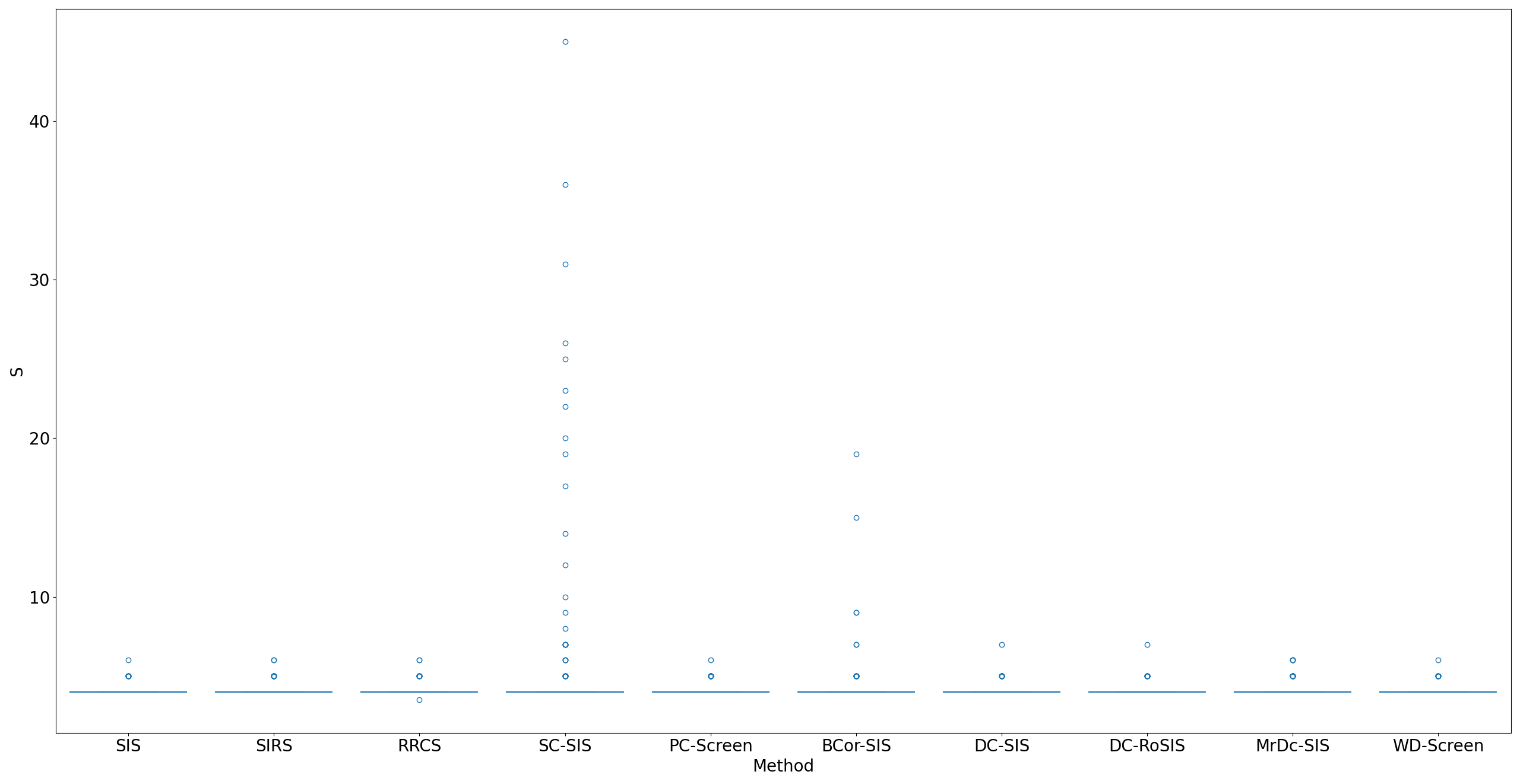}
\caption{$\mathit{S}$ in Study 1. The smaller the $\mathit{S}$, the better the performance.}
\end{minipage}
\hfil
\begin{minipage}{.48\textwidth}
\includegraphics[width=\textwidth]{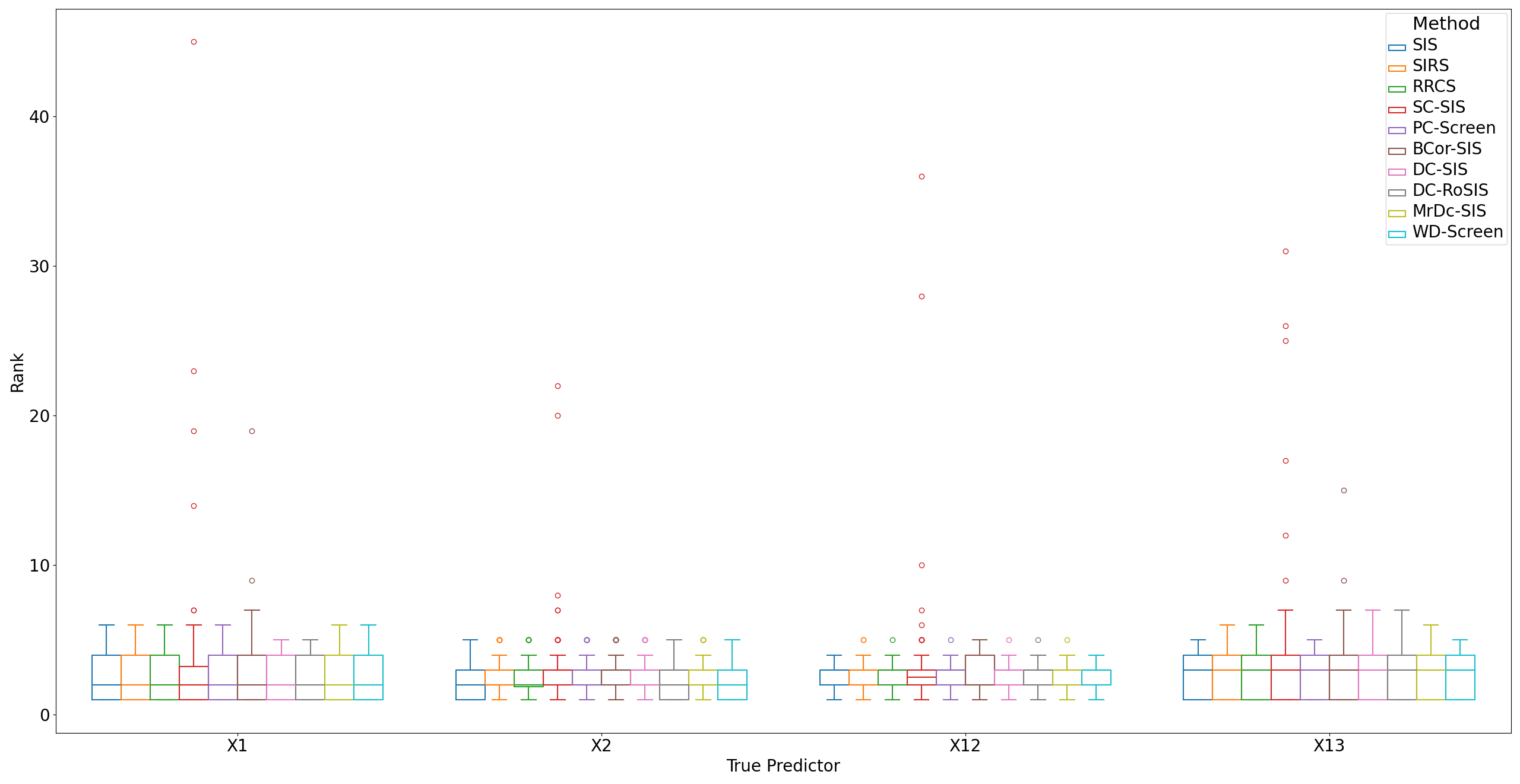}
\caption{Rank of true predictors in Study 1. The smaller the rank, the better the performance.}
\end{minipage}
\label{simple_normal}
\end{figure}

\textcolor{red}{We can see all methods performed pretty well in this simple settings.}

\begin{table}[h]
\begin{center}
\caption{Performance of the three approaches for Study 1. The individual success rate $\mathit{P}_s$, and the overall success rate $\mathit{P}_a$ are demonstrated. The predetermined cutoff for $\mathit{P}_s$ and $\mathit{P}_a$ is $s = [n/\log(n)]=37$.}
\label{tblstudy1}
\footnotesize
\begin{tabular}{ccccccccccccc}
\toprule
Method  & $\mathit{P}_{\Xv_1}$ & $\mathit{P}_{\Xv_2}$ & $\mathit{P}_{\Xv_{12}}$ & $\mathit{P}_{\Xv_{13}}$ & $\mathit{P}_a$ & & Method  & $\mathit{P}_{\Xv_1}$ & $\mathit{P}_{\Xv_2}$ & $\mathit{P}_{\Xv_{12}}$ & $\mathit{P}_{\Xv_{13}}$ & $\mathit{P}_a$ \\
    \midrule
SIS & 1 & 1 & 1 & 1  & 1 & \vline & BCor-SIS & 1 & 1 & 1 & 1  & 1 \\

SIRS & 1 & 1 & 1 & 1  & 1 & \vline & DC-SIS & 1 & 1 & 1 & 1  & 1 \\

RRCS & 1 & 1 & 1 & 1  & 1 &\vline &  DC-RoSIS & 1 & 1 & 1 & 1  & 1 \\

SC-SIS & 0.995 & 1 & 1 & 1  & 0.995 & \vline & MrDc-SIS & 1 & 1 & 1 & 1  & 1 \\

PC-Screen & 1 & 1 & 1 & 1  & 1 & \vline & WD-Screen & 1 & 1 & 1 & 1  & 1 \\
 \bottomrule
\end{tabular}
\end{center}
\end{table}

\subsubsection*{Study 2: Evaluating Performance with Non-Normal Predictor Distributions}
In this study, we change the distribution of the predictors, $\Xv_i$, to follow an i.i.d. power function distribution, which is the inverse of the Pareto distribution with probability density function $f(x; a) = a x^{a-1}$ for $0< x< 1$ and $a> 0$. In our simulation, we choose the parameter $a = 5$, A typical histogram of such distribution below illustrates how it more closely resembles real-world data than the normal distribution used in Study 1. To further increase the difficulty of the test, the coefficients $\beta_{1,2,3,4}\sim Uniform (1,2)$, which is smaller than in Study 1. All other settings remain the same as in Study 1.

\begin{figure}[h!]
\centering
\begin{minipage}{.48\textwidth}
\includegraphics[width=\textwidth]{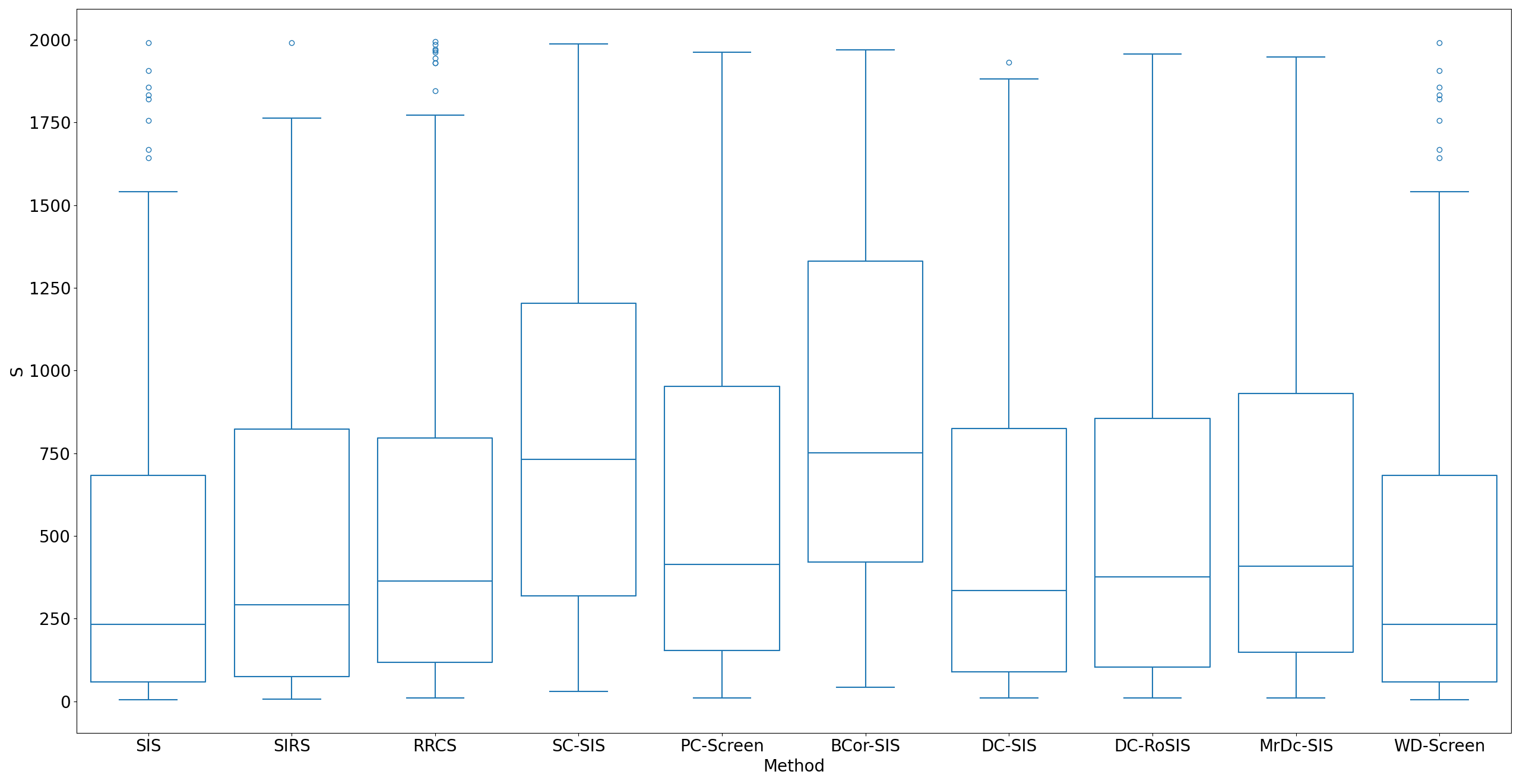}
\caption{$\mathit{S}$ in Study 2. The smaller the $\mathit{S}$, the better the performance.}
\end{minipage}
\hfil
\begin{minipage}{.48\textwidth}
\includegraphics[width=\textwidth]{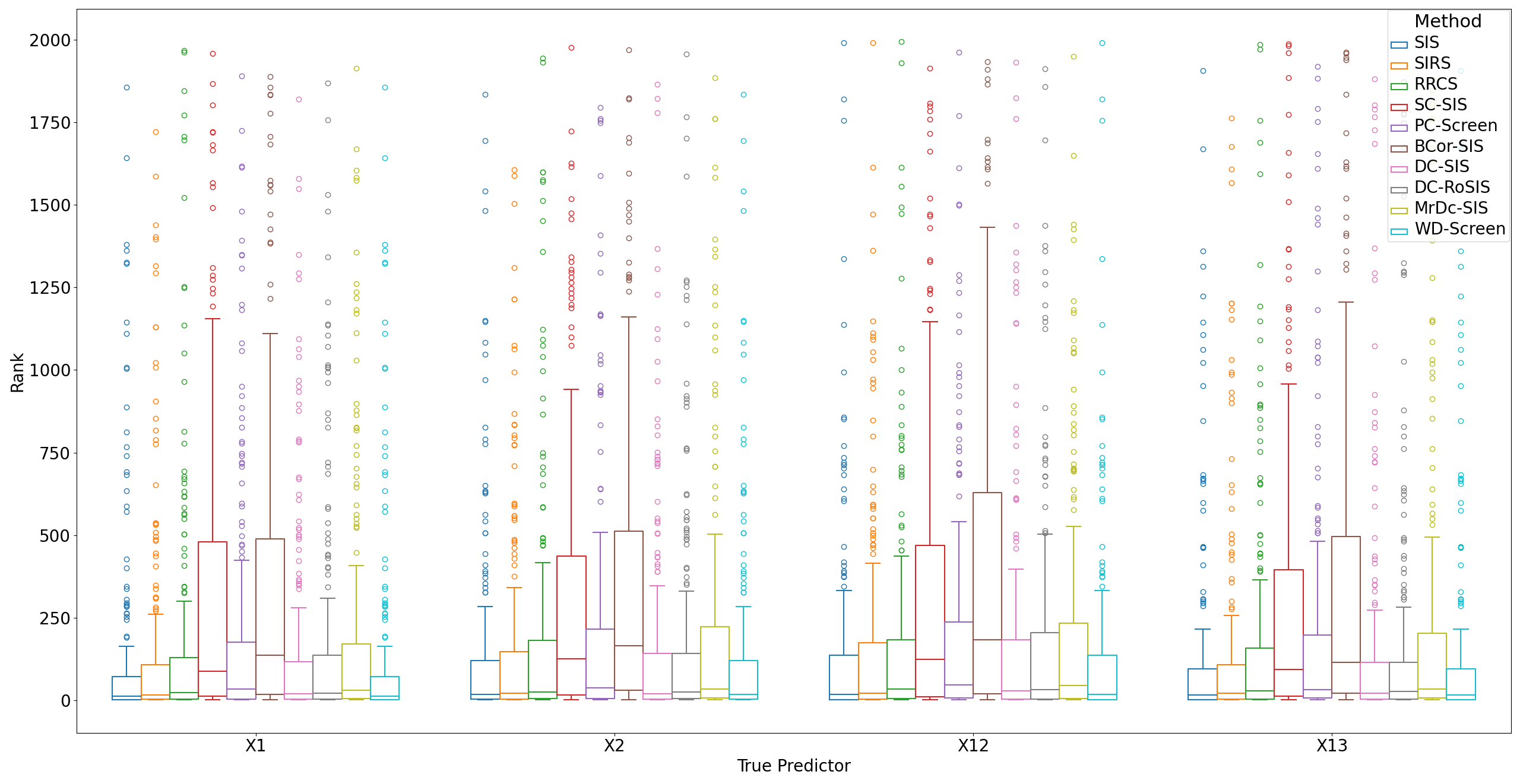}
\caption{Rank of true predictors in Study 2. The smaller the rank, the better the performance.}
\end{minipage}
\label{single_power}
\end{figure}

\textcolor{red}{We can see if the predictors are from power distribution, the performance of all methods get worse. But still the WD-Screen performs better than other methods.}

\begin{table}[h]
\begin{center}
\caption{Performance of the three approaches for Study 1. The individual success rate $\mathit{P}_s$, and the overall success rate $\mathit{P}_a$ are demonstrated. The predetermined cutoff for $\mathit{P}_s$ and $\mathit{P}_a$ is $s = [n/\log(n)]=37$.}
\label{tblstudy2}
\footnotesize
\begin{tabular}{ccccccccccccc}
\toprule
Method  & $\mathit{P}_{\Xv_1}$ & $\mathit{P}_{\Xv_2}$ & $\mathit{P}_{\Xv_{12}}$ & $\mathit{P}_{\Xv_{13}}$ & $\mathit{P}_a$ & & Method  & $\mathit{P}_{\Xv_1}$ & $\mathit{P}_{\Xv_2}$ & $\mathit{P}_{\Xv_{12}}$ & $\mathit{P}_{\Xv_{13}}$ & $\mathit{P}_a$ \\
    \midrule
SIS & 0.66 & 0.605 & 0.6 & 0.62  & 0.165 & \vline & BCor-SIS & 0.335 & 0.295 & 0.3 & 0.345 & 0 \\

SIRS & 0.61 & 0.59 & 0.57 & 0.61  & 0.14 & \vline & DC-SIS & 0.615 & 0.57 & 0.535 & 0.585  & 0.11 \\

RRCS & 0.595 & 0.56 & 0.53 & 0.55  & 0.1 &\vline &  DC-RoSIS & 0.585 & 0.565 & 0.53 & 0.59  & 0.11 \\

SC-SIS & 0.35 & 0.33 & 0.36 & 0.385 & 0.005 & \vline & MrDc-SIS & 0.53 & 0.515 & 0.47 & 0.52 & 0.05 \\

PC-Screen & 0.53 & 0.485 & 0.465 & 0.54 & 0.055 & \vline & WD-Screen & 0.66 & 0.605 & 0.6 & 0.62  & 0.165 \\
 \bottomrule
\end{tabular}
\end{center}
\end{table}

\subsubsection*{Study 3: Simulating Multi-Omics and Multi-Endpoint Data Structures}
To more closely simulate the structure of multi-omics data, we adapted a settings based on a TCGA-LUAD simulation from Zhao \textit{et al.}\cite{zhao2024model} In this study, we use again $n= 200$ observations, with $p = 2000$ predictor array and a $q = 10$-dimensional response vector. Each predictor is represented as a 3-dimensional vector to mimic the multi-omics settings, where data are collected across 3 different platforms (e.g. copy number variation, RNAseq, etc.). The 10-dimensional response vector, in turn, reflects real-world situations where multiple clinical or biological endpoints are analyzed simultaneously.

The 3 platforms are generated as follows:
\begin{itemize}
    \item Platform 1 ($\Uv = [U_1, U_2, \dots, U_p]$): Multivariate Pareto distributed with shape $a_U = 10$ and mode $m_U = 1$.
    \item Platform 2 ($\Vv = [V_1, V_2, \dots, V_p]$): Multivariate Power distributed with parameter $a_V = 5$.
    \item Platform 3 ($\Wv = [W_1, W_2, \dots, W_p]$): Multivariate Power distributed with parameter $a_W = 5$.
\end{itemize}
Each platform has a shared covariance structure $\Sigmav_{p\times p} = [\sigma_{ij}]_{p\times p}$, where $\sigma_{ij} = 0.5^{|i-j|}$, and the predictor array $\Xbf_{3\times n\times p} = [\Xv_1, \Xv_2, \dots, \Xv_p]$ is constructed by stacking these platforms: $\Xv_j = [\Uv_j, \Vv_j, \Wv_j]$, $\forall j = 1, 2, \dots, p$.

The response vector is connected with the predictors as follows:
\begin{itemize}
    \item For $k = 1,2,3$
    \begin{enumerate}
        \item Randomly select indices $id_{1,2,3,4}$ from $\{1,2,3\}$ to represent the true platforms connected with the response.
        \item $\Yv_k[i] = \beta_1 \Xbf_{id_1, i, 2} + \beta_2 \Xbf_{id_2, i, 3} + \beta_3 \Xbf_{id_3, i, 101} + \beta_4 \Xbf_{id_4, i, 102} + \epsilonv[i]$, $\forall i = 1,2, \dots, n$, where $\beta_{1,2,3,4}\sim Uniform (1,2)$ and $\epsilonv\sim N(0, 1)$.
    \end{enumerate}
    \item For $k = 4, 5, \dots, 10$, $\Yv_k\sim Power (5)$.
\end{itemize}
Since SIS, SIRS, RRCS, DC-RoSIS cannot handle multivariate response and predictors, we only compare the performance of other 6 methods.

\begin{figure}[h!]
\centering
\begin{minipage}{.48\textwidth}
\includegraphics[width=\textwidth]{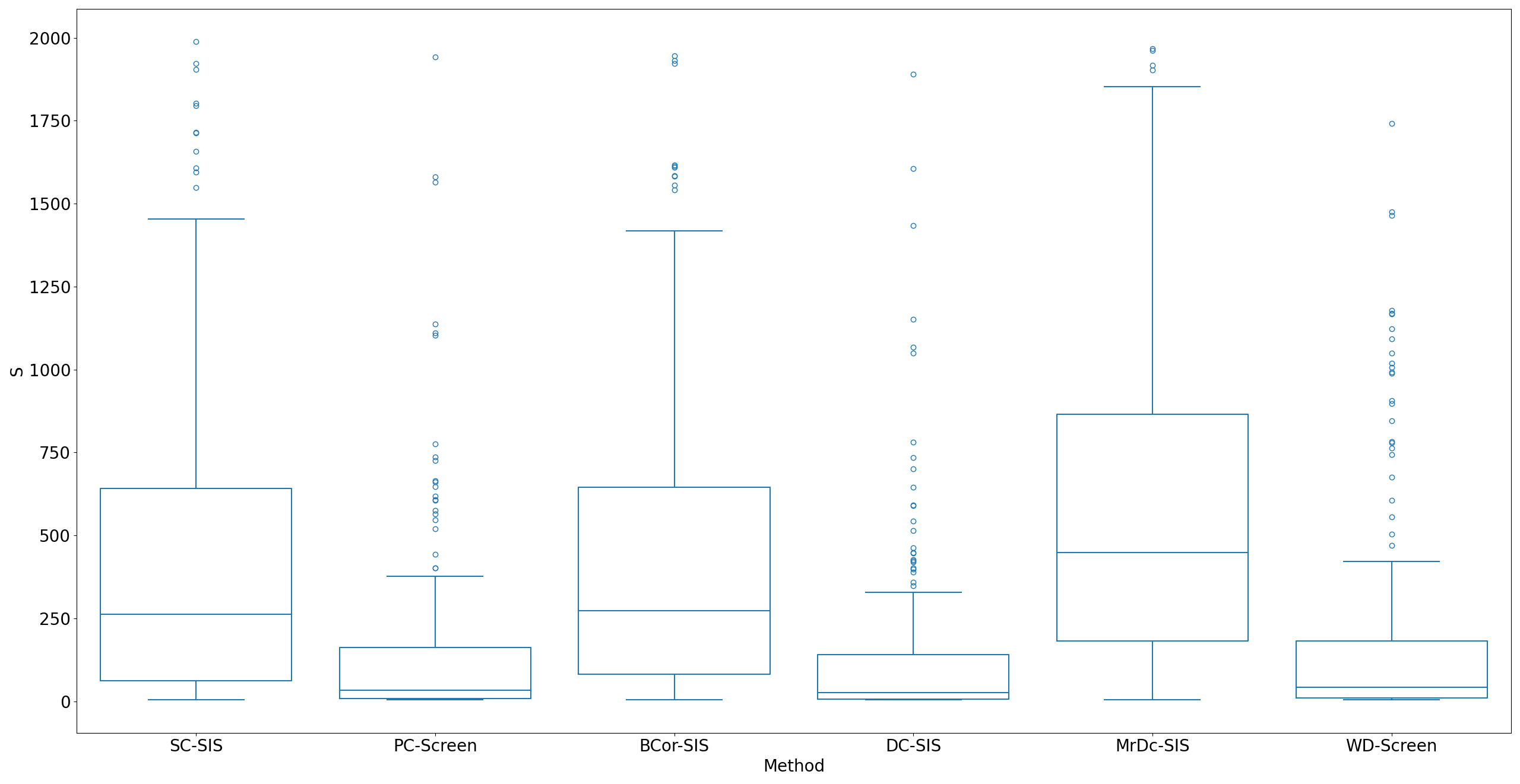}
\caption{$\mathit{S}$ in Study 3. The smaller the $\mathit{S}$, the better the performance.}
\end{minipage}
\hfil
\begin{minipage}{.48\textwidth}
\includegraphics[width=\textwidth]{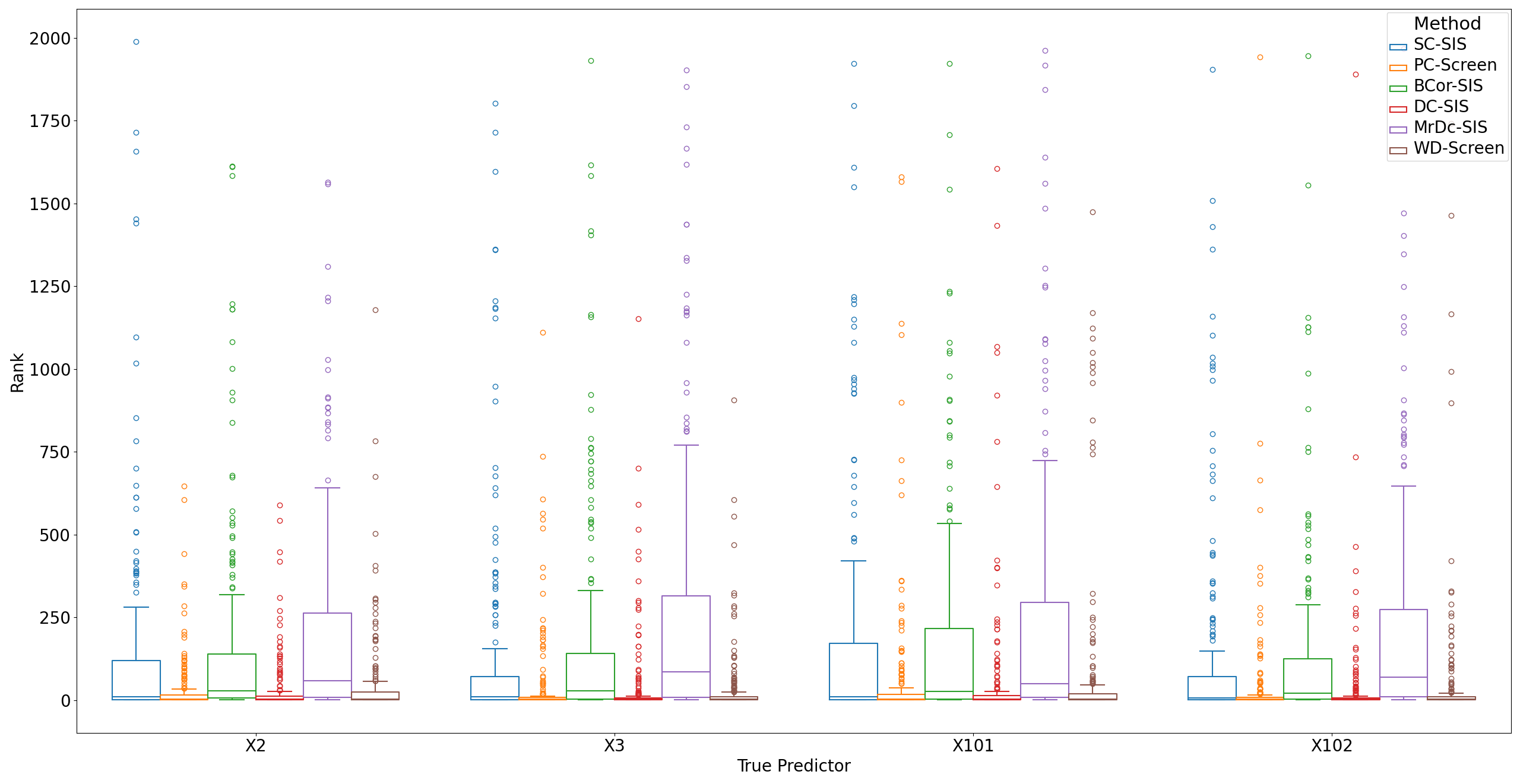}
\caption{Rank of true predictors in Study 3. The smaller the rank, the better the performance.}
\end{minipage}
\label{multi_power}
\end{figure}

\textcolor{red}{We can see clearly under this settings, PC-Screen, DC-SIS, and WD-Screen works constantly better than others.}
\begin{table}[h]
\begin{center}
\caption{Performance of the three approaches for Study 1. The individual success rate $\mathit{P}_s$, and the overall success rate $\mathit{P}_a$ are demonstrated. The predetermined cutoff for $\mathit{P}_s$ and $\mathit{P}_a$ is $s = [n/\log(n)]=37$.}
\label{tblstudy3}
\footnotesize
\begin{tabular}{cccccc}
\toprule
Method  & $\mathit{P}_{\Xv_2}$ & $\mathit{P}_{\Xv_3}$ & $\mathit{P}_{\Xv_{101}}$ & $\mathit{P}_{\Xv_{102}}$ & $\mathit{P}_a$  \\
    \midrule
SC-SIS & 0.595 & 0.68 & 0.65 & 0.675  & 0.195 \\
PC-Screen & 0.83 & 0.85 & 0.815 & 0.865  & 0.505 \\
BCor-SIS & 0.575 & 0.56 & 0.545 & 0.6  & 0.14  \\
DC-SIS & 0.85 & 0.86 & 0.835 & 0.87 & 0.555  \\
MrDc-SIS & 0.43 & 0.4 & 0.47 & 0.4 & 0.035  \\
WD-Screen & 0.8 & 0.83 & 0.8 & 0.845 & 0.475 \\
 \bottomrule
\end{tabular}
\end{center}
\end{table}

\subsubsection*{Study 4: Assessing Feature Selection with Interaction Effects}
In this study, we introduce an interaction term to simulate real-world scenarios where certain genes may not have an individual effect on the disease, but may influence it in combination with others. Most settings remain the same as in Study 3, but the active response for $k = 1,2,3$ is now
\[
\Yv_k[i] = \beta_1 \Xbf_{id_1, i, 2} \times \Xbf_{id_2, i, 3} + \beta_2 \Xbf_{id_3, i, 101} \times  \Xbf_{id_4, i, 102} + \epsilonv[i], \forall i = 1,2, \dots,n,
\]
where $\beta_{1,2}\sim Uniform (1,2)$ and $\epsilonv\sim N(0, 1)$.
\begin{figure}[h]
\centering
\begin{minipage}{.48\textwidth}
\includegraphics[width=\textwidth]{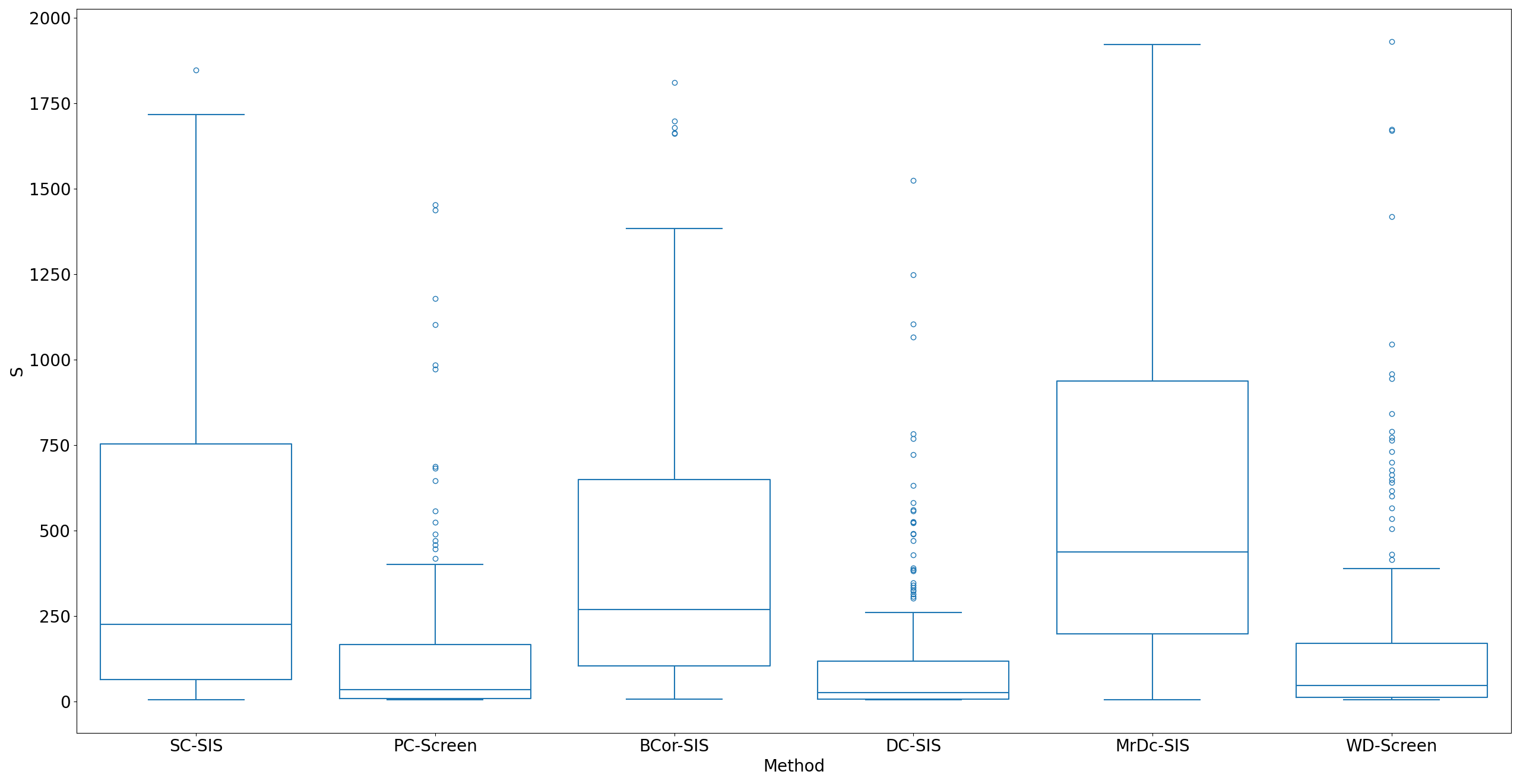}
\caption{$\mathit{S}$ in Study 4. The smaller the $\mathit{S}$, the better the performance.}
\end{minipage}
\hfil
\begin{minipage}{.48\textwidth}
\includegraphics[width=\textwidth]{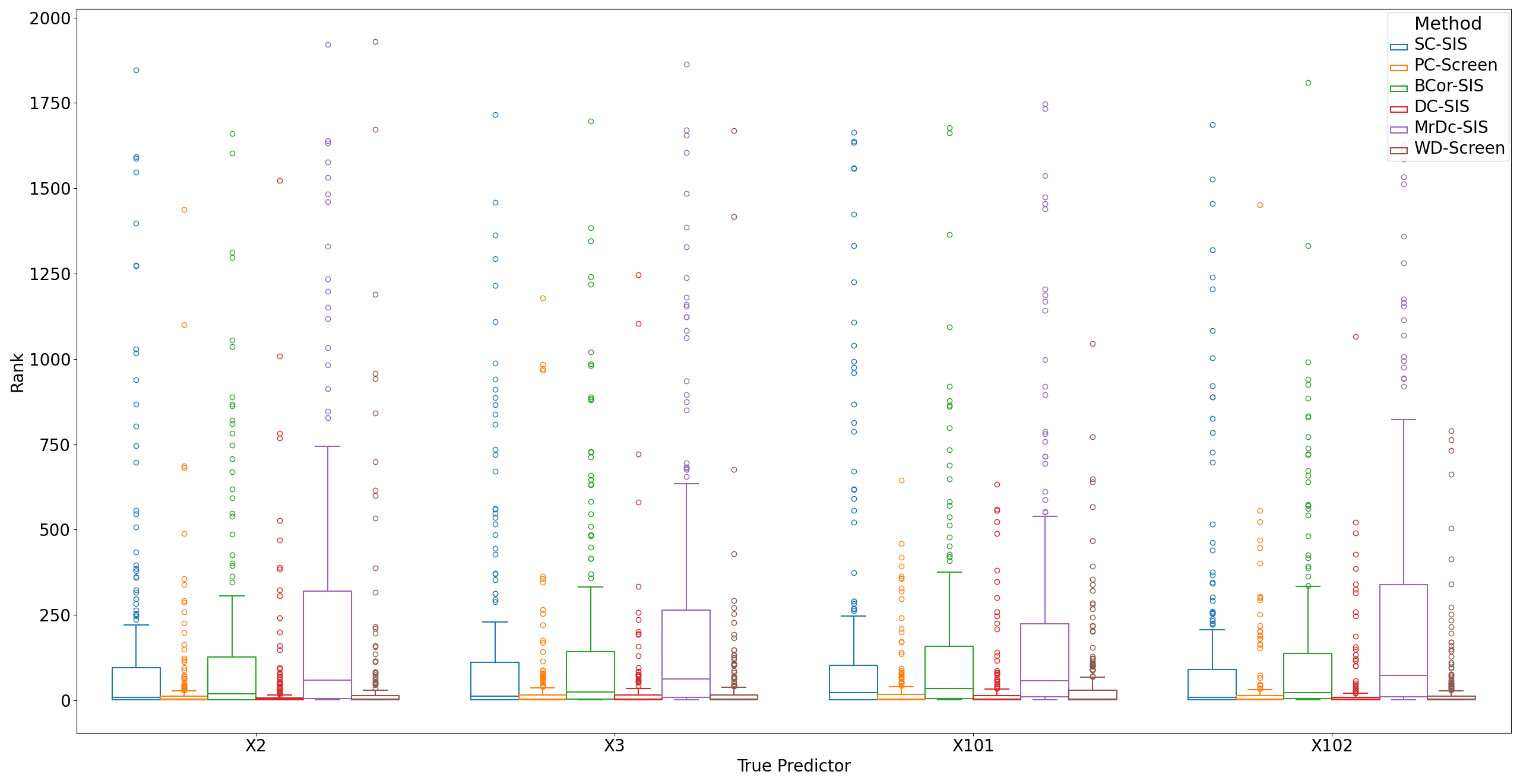}
\caption{Rank of true predictors in Study 4. The smaller the rank, the better the performance.}
\end{minipage}
\label{multi_power_interaction}
\end{figure}

\textcolor{red}{We can see again PC-Screen, DC-SIS, and WD-Screen works better than other methods.}
\begin{table}[h]
\begin{center}
\caption{Performance of the three approaches for Study 1. The individual success rate $\mathit{P}_s$, and the overall success rate $\mathit{P}_a$ are demonstrated. The predetermined cutoff for $\mathit{P}_s$ and $\mathit{P}_a$ is $s = [n/\log(n)]=37$.}
\label{tblstudy4}
\footnotesize
\begin{tabular}{cccccc}
\toprule
Method  & $\mathit{P}_{\Xv_2}$ & $\mathit{P}_{\Xv_3}$ & $\mathit{P}_{\Xv_{101}}$ & $\mathit{P}_{\Xv_{102}}$ & $\mathit{P}_a$  \\
    \midrule
SC-SIS & 0.63 & 0.64 & 0.58 & 0.655  & 0.17 \\
PC-Screen & 0.83 & 0.82 & 0.825 & 0.88  & 0.51 \\
BCor-SIS & 0.58 & 0.57 & 0.52 & 0.55  & 0.09  \\
DC-SIS & 0.855 & 0.87 & 0.865 & 0.89 & 0.575  \\
MrDc-SIS & 0.415 & 0.405 & 0.46 & 0.395 & 0.035  \\
WD-Screen & 0.835 & 0.84 & 0.76 & 0.81 & 0.44 \\
 \bottomrule
\end{tabular}
\end{center}
\end{table}

\subsection*{Feature Selection Methods on Real-World Data: A TCGA-LUAD Case Study}
We obtained the TCGA-LUAD data from cBioPortal\cite{cerami2012cbio} and selected the nonsynonymous Tumor Mutational Burden (TMB) from clinical sample data as our response variable. TMB, particularly nonsynonymous TMB, is an important biomarker in cancer research, measuring the total number of somatic nonsynonymous mutations per megabase in tumor cells, and it can vary widely across and within cancer types. High TMB levels increase the production of neoantigens, which may be recognized by the immune system, potentially enhancing the efficacy of immunotherapy. Recently, studies have shown that TMB is associated with clinical outcomes in multiple cancers, including melanoma, non-small-cell lung cancer, and colorectal cancer. Evidence suggests that high TMB can effectively predict objective response rates and progression-free survival, making it a valuable indicator in assessing immunotherapy outcomes.\cite{budczies2024tumour, li2020choosing}.

In our study, we treat TMB as the response variable and seek to understand its association with genetic variations by integrating data from multi-omics platforms. Specifically, we use data on copy number alteration (CNA), DNA methylation, and mRNA expression as predictors. Each of these platforms provides unique insights: CNA data reflect gene amplification or deletion, methylation data reveal epigenetic changes that may influence gene expression, and mRNA expression levels indicate gene activity. By combining information from these platforms, we aim to identify genes whose variations correlate strongly with TMB, uncovering potential drivers of mutational burden in LUAD. This multi-omics approach allows us to explore complex, cross-platform interactions and their influence on TMB, which could ultimately inform targeted strategies for immunotherapy in lung cancer.

From our simulation studies, we observe that PC-Screen, DC-SIS, and WD-Screen consistently outperform other feature selection methods, showing reliable results across diverse scenarios. For the real data analysis, we apply these 3 methods to our dataset, each offering a distinct approach and methodology to feature selection. Instead of using a traditional training-test split, we adopt a more robust selection strategy which can use the full information of the real data: for each method, we select the top $[n/\log(n)]$ features to form a selection set, then choose the intersection of the 3 selection sets as our final selection set. By focusing on features that are independently identified by all 3 top-performing methods, we aim to enhance the robustness and reliability of our final selection, ensuring that the chosen features are more likely to have genuine associations with TMB.

\subsubsection*{2-platform study}
In our first study, we combine CNA and mRNA data, using the files ``data\_mrna\_seq\_v2\_rsem\_zscores\_ref\_all\_samples.txt'' for mRNA and ``data\_cna.txt'' for CNA. The original CNA data contains 230 subjects with 23423 genes, while the mRNA data contains 230 subjects with 20466 genes. After removing the duplicates and genes missing from one platform, there are 18674 common genes across 230 subjects in these 2 platforms. Our predictor $\Xbf$ will be a $2\times 230\times 18674$ array, and our response $\Yv$ will be a $230\times 1$ vector.

The 3 methods have 13 genes in common among the top $[230/\log(230)] = 42$ selected genes: \emph{CCNG1, CKAP2L, DTWD2, FLJ33630, HSD17B4, NME5, NUDT12, PCBD2, REEP5, SHROOM1, SLC22A5, TIGD6}, and \emph{TMEM173}. Below is a visulization of the relationship between TMB and the CNA and mRNA of the 13 genes, the results also show that these 3 methods can handle categorical variables and continuous variables together in one predictor vector. 

\begin{figure}[h]
\centering
\includegraphics[width=\textwidth]{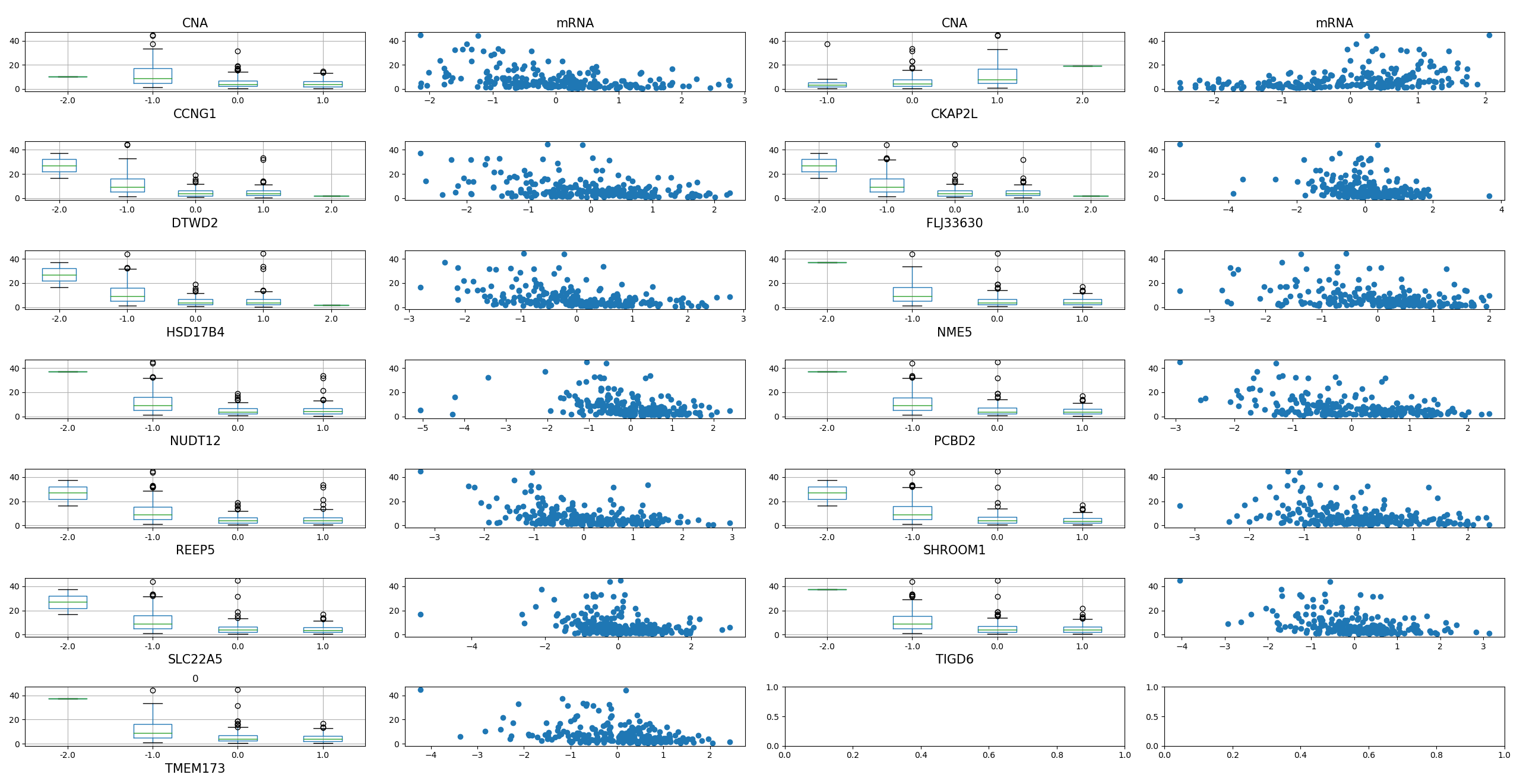}
\caption{The CNA vs TMB, and mRNA vs TMB plots of the 13 selected genes.}
\label{2plat}
\end{figure}

\subsubsection*{3-platform study}
14399 genes in 185 subjects,
'CCNG1', 'DTWD2', 'PCBD2', 'TMEM173' : top 43 has 4 common
'PCBD2', 'TMEM173' top 36 has 2 common

\section*{Discussion}

The Discussion should be succinct and must not contain subheadings.

\section*{Methods}

Topical subheadings are allowed. Authors must ensure that their Methods section includes adequate experimental and characterization data necessary for others in the field to reproduce their work.

\bibliography{sample}

@MISC{GlobalCancer,
    author = {Ferlay J and Ervik M and Lam F and Laversanne M and Colombet M and Mery L and Piñeros M and Znaor A and Soerjomataram I and Bray F},
    year = {2024},
    title = {Global Cancer Observatory: Cancer Today},
    howpublished = {\url{https://gco.iarc.who.int/today}},
    note = {Accessed 28 Oct 2024}
}

@article{abdelwahab2022feature,
  title={A feature selection-based framework to identify biomarkers for cancer diagnosis: A focus on lung adenocarcinoma},
  author={Abdelwahab, Omar and Awad, Nourelislam and Elserafy, Menattallah and Badr, Eman},
  journal={Plos one},
  volume={17},
  number={9},
  pages={e0269126},
  year={2022},
  publisher={Public Library of Science San Francisco, CA USA}
}

@article{zhao2024model,
  title={A model-free and distribution-free multi-omics integration approach for detecting novel lung adenocarcinoma genes},
  author={Zhao, Shaofei and Qi, Caleb and Zhao, Geran and Wang, Yangsheng and Fu, Guifang},
  journal={Scientific Reports},
  volume={14},
  number={1},
  pages={17996},
  year={2024},
  publisher={Nature Publishing Group UK London}
}

@article{fan2008sure,
  title={Sure independence screening for ultrahigh dimensional feature space},
  author={Fan, Jianqing and Lv, Jinchi},
  journal={Journal of the Royal Statistical Society Series B: Statistical Methodology},
  volume={70},
  number={5},
  pages={849--911},
  year={2008},
  publisher={Oxford University Press}
}

@article{zhao2022distribution,
  title={Distribution-free and model-free multivariate feature screening via multivariate rank distance correlation},
  author={Zhao, Shaofei and Fu, Guifang},
  journal={Journal of Multivariate Analysis},
  volume={192},
  pages={105081},
  year={2022},
  publisher={Elsevier}
}

@article{li2012feature,
  title={Feature screening via distance correlation learning},
  author={Li, Runze and Zhong, Wei and Zhu, Liping},
  journal={Journal of the American Statistical Association},
  volume={107},
  number={499},
  pages={1129--1139},
  year={2012},
  publisher={Taylor \& Francis}
}

@article{guo2022stable,
  title={Stable correlation and robust feature screening},
  author={Guo, Xu and Li, Runze and Liu, Wanjun and Zhu, Lixing},
  journal={Science China Mathematics},
  volume={65},
  number={1},
  pages={153--168},
  year={2022},
  publisher={Springer}
}

@article{mordant2022measuring,
  title={Measuring dependence between random vectors via optimal transport},
  author={Mordant, Gilles and Segers, Johan},
  journal={Journal of Multivariate Analysis},
  volume={189},
  pages={104912},
  year={2022},
  publisher={Elsevier}
}

@article{cerami2012cbio,
  title={The cBio cancer genomics portal: an open platform for exploring multidimensional cancer genomics data},
  author={Cerami, Ethan and Gao, Jianjiong and Dogrusoz, Ugur and Gross, Benjamin E and Sumer, Selcuk Onur and Aksoy, B{\"u}lent Arman and Jacobsen, Anders and Byrne, Caitlin J and Heuer, Michael L and Larsson, Erik and others},
  journal={Cancer discovery},
  volume={2},
  number={5},
  pages={401--404},
  year={2012},
  publisher={AACR}
}

@article{zhong2016regularized,
  title={Regularized quantile regression and robust feature screening for single index models},
  author={Zhong, Wei and Zhu, Liping and Li, Runze and Cui, Hengjian},
  journal={Statistica Sinica},
  volume={26},
  number={1},
  pages={69},
  year={2016},
  publisher={NIH Public Access}
}

@article{zhu2011model,
  title={Model-free feature screening for ultrahigh-dimensional data},
  author={Zhu, Li-Ping and Li, Lexin and Li, Runze and Zhu, Li-Xing},
  journal={Journal of the American Statistical Association},
  volume={106},
  number={496},
  pages={1464--1475},
  year={2011},
  publisher={Taylor \& Francis}
}

@article{li2012robust,
  title={Robust rank correlation based screening},
  author={Li, Gaorong and Peng, Heng and Zhang, Jun and Zhu, Lixing},
  journal={The Annals of Statistics},
  volume={40},
  number={3},
  pages={1846--1877},
  year={2012},
  publisher={Institute of Mathematical Statistics}
}

@article{liu2020model,
  title={Model-free feature screening and {FDR} control with {K}nockoff features},
  author={Liu, Wanjun and Ke, Yuan and Liu, Jingyuan and Li, Runze},
  journal={Journal of the American Statistical Association},
  pages={1--16},
  year={2020},
  publisher={Taylor \& Francis}
}

@article{pan2018generic,
  title={A generic sure independence screening procedure},
  author={Pan, Wenliang and Wang, Xueqin and Xiao, Weinan and Zhu, Hongtu},
  journal={Journal of the American Statistical Association},
  year={2018},
  publisher={Taylor \& Francis}
}

@article{budczies2024tumour,
  title={Tumour mutational burden: Clinical utility, challenges and emerging improvements},
  author={Budczies, Jan and Kazdal, Daniel and Menzel, Michael and Beck, Susanne and Kluck, Klaus and Altb{\"u}rger, Christian and Schwab, Constantin and Allg{\"a}uer, Michael and Ahadova, Aysel and Kloor, Matthias and others},
  journal={Nature Reviews Clinical Oncology},
  pages={1--18},
  year={2024},
  publisher={Nature Publishing Group UK London}
}

@article{li2020choosing,
  title={Choosing tumor mutational burden wisely for immunotherapy: a hard road to explore},
  author={Li, Rui and Han, Dongsheng and Shi, Jiping and Han, YanXi and Tan, Ping and Zhang, Rui and Li, Jinming},
  journal={Biochimica et Biophysica Acta (BBA)-Reviews on Cancer},
  volume={1874},
  number={2},
  pages={188420},
  year={2020},
  publisher={Elsevier}
}

For data citations of datasets uploaded to e.g. \emph{figshare}, please use the \verb|howpublished| option in the bib entry to specify the platform and the link, as in the \verb|Hao:gidmaps:2014| example in the sample bibliography file.

\section*{Acknowledgements (not compulsory)}

Acknowledgements should be brief, and should not include thanks to anonymous referees and editors, or effusive comments. Grant or contribution numbers may be acknowledged.

\section*{Author contributions statement}

Must include all authors, identified by initials, for example:
A.A. conceived the experiment(s),  A.A. and B.A. conducted the experiment(s), C.A. and D.A. analysed the results.  All authors reviewed the manuscript. 

\section*{Additional information}

To include, in this order: \textbf{Accession codes} (where applicable); \textbf{Competing interests} (mandatory statement). 

The corresponding author is responsible for submitting a \href{http://www.nature.com/srep/policies/index.html#competing}{competing interests statement} on behalf of all authors of the paper. This statement must be included in the submitted article file.

\end{document}